# Legal Challenges in Renewable Energy Development: A Comparative Study of China and Selected Countries

Aliasghar Baziar, *Member, IEEE*, Navid Parsa, *Member, IEEE*

*Authors are with Electrical and Computer Engineering Department, Shiraz University of Technology, Shiraz, Fars, Iran*

**Abstract:** This exhaustive investigation is dedicated to delving into the intricate legal aspects that underlie the inefficiency in the advancement and utilization of sustainable energies, with a primary focus on the dynamic landscape of China and carefully selected representative nations. In an era where the global community increasingly acknowledges the pressing need for environmentally-friendly alternatives to traditional fossil fuels, renewable energy sources have rightfully garnered substantial attention as encouraging solutions. Nevertheless, notwithstanding their potential to revolutionize the energy sector and counteract climate change, a multitude of legal and regulatory barriers may present formidable hindrances that impede their seamless integration into the energy landscape. With a resolute and concentrated aim, the research sets forth on a painstaking exploration and analysis of the intricate legal frameworks, policies, and institutional arrangements in place within China and the chosen representative nations. The ultimate objective is to discern and identify potential challenges and inefficiencies that could hinder the progress of renewable energy projects and initiatives. By undertaking a stringent and in-depth examination, this study endeavors to illuminate the specific barriers that hinder the effective implementation of renewable energy endeavors. Employing a comparative approach, the research delves into unraveling the ways distinct legal systems influence the development and application of renewable energies in each country under scrutiny. This comparative perspective provides a unique opportunity to gain a nuanced understanding of how various legal environments can either promote or obstruct the growth of renewable energy industries. Furthermore, the analysis extends to evaluating the effectiveness of various incentives, subsidies, and environmental regulations, which play a crucial role in promoting and facilitating the widespread adoption of renewable energy technologies. Moreover, this research goes above and beyond by delving deeply into enlightening case studies and extracting best practices from carefully chosen countries that have demonstrated successful methods of overcoming legal obstacles in renewable energy development. These case studies offer invaluable insights and valuable lessons that can guide policymakers, legislators, and industry stakeholders in devising effective strategies to navigate the complex and ever-evolving legal landscape surrounding renewable energies. The implications of this study's findings hold vast potential for the global advancement of renewable energy adoption. By identifying areas for enhancement and recognizing the key elements that drive successful integration, this research aims to play a pivotal role in shaping and influencing the transition towards a more sustainable and eco-friendly future. As societies worldwide grapple with the urgent need to reshape their energy landscapes for the better, the outcomes of this study have the potential to pave the way for a greener, cleaner, and more resilient world. In a world where the consequences of climate change are becoming increasingly conspicuous, the urgency to embrace renewable energies has never been more significant. This study's comprehensive analysis of legal barriers and regulatory challenges seeks to empower decision-makers and stakeholders with the knowledge and insights necessary to promote an environment conducive to renewable energy development. By fostering a profound comprehension of the legal landscape, this research aims to propel progressive policy reforms, bolster effective governance, and inspire innovative approaches to hasten the adoption and integration of renewable energies on a global scale. As the urgency to combat climate change continues to intensify, the transformative potential of renewable energies cannot be underestimated. This study's contributions are envisioned to act as a catalyst for a greener and more sustainable world, where renewable energies stand at the forefront of a resilient and low-carbon energy future. By bridging the gap between legal intricacies and renewable energy development, this research seeks to ignite positive change and contribute to the collective endeavors to construct a better, cleaner, and more sustainable tomorrow.

1. **Introduction**

The globe stands at a crucial intersection, wrestling with the burgeoning apparition of climate change and its profound ramifications for the future of the planet [1-3]. As communities acknowledge the pressing demand for sustainable energy solutions, sustainable energies have emerged as a guiding light, offering the assurance of a more ecologically aware and verdant future [4]. The escalating realization of the harmful repercussions of conventional fossil fuels on the ecosystem has led to a growing universal agreement on the necessity to shift towards renewable energy sources [5-8]. Nevertheless, this shift is not without its difficulties. Notwithstanding the potential of renewable energy technologies to revolutionize the energy sector and alleviate the undesirable consequences of climate change, a plethora of legal and regulatory hindrances have emerged, obstructing their effective advancement and seamless incorporation into the energy landscape [9-12].

Amidst this setting, this research initiates a crucial and all-encompassing mission to investigate and scrutinize the elaborate legal facets that underpin the inefficacy in the growth and utilization of renewable energies, with a specific emphasis on China and meticulously chosen representative nations [13]. The importance of this exploration cannot be overemphasized, as the efficient utilization and extensive implementation of renewable energy technologies call for traversing an intricate labyrinth of legal structures, policies, and institutional setups, distinct to the context of each country [14-17].

The fundamental aim of this investigation is to pinpoint and assess the legal hurdles and shortcomings that hinder the smooth advancement of renewable energy projects and endeavors in the scrutinized countries [18-20]. With steadfast resolve, this study seeks to illuminate the particular obstacles that obstruct the efficient incorporation and utilization of renewable energies in the energy framework of these nations [21]. The implementation of a comparative approach serves as a defining characteristic of this study, providing a unique chance to grasp the varying consequences of diverse legal systems on the progress of renewable energy in distinct national settings [22-23]. Through meticulous examination of the legal frameworks and regulatory landscapes in China and other representative countries, this research aims to unravel the ways in which disparities in legal systems can either foster or hinder the growth and advancement of renewable energy sectors [24-26].

Furthermore, this investigation seeks to evaluate the effectiveness of different encouragements, subsidies, and environmental regulations in stimulating and facilitating the extensive embrace of renewable energy technologies. Through a comprehensive analysis, this study aims to shed light on the merits and shortcomings of current legal mechanisms in promoting the shift towards renewable energies. Building on its dedication, this study delves extensively into the domain of elucidating case studies and extracting exemplary approaches from meticulously selected countries that have effectively overcome legal impediments in their pursuit of renewable energy development. These case studies offer priceless perspectives and teachings that can steer policymakers, legislators, and industry participants in formulating potent strategies to navigate the intricate and continuously evolving legal terrain related to renewable energies [27].

The ramifications of the conclusions derived from this study reverberate on a planetary scale, carrying the capacity to profoundly contribute to the global progress of adopting renewable energies. Through pinpointing precise domains for enhancement and acknowledging the pivotal factors that foster successful assimilation, this research endeavors to assume a crucial position in shaping and influencing the path towards a more sustainable and environmentally friendly future. As the urgency to confront the repercussions of climate change intensifies, embracing renewable energies has acquired a position of utmost significance in preserving the ecological equilibrium of our planet. This study's meticulous scrutiny of legal aspects and regulatory hurdles is positioned to empower decision-makers and stakeholders with the knowledge and insights indispensable for fostering an environment conducive to the development of renewable energies. By bridging the divide between legal complexities and the pragmatic realization of renewable energy projects, this research endeavors to catalyze positive transformation and contribute to the collective global endeavor to construct a better, cleaner, and more sustainable tomorrow. As we navigate the path ahead, characterized by unparalleled environmental challenges, this study envisages a world where renewable energies steadfastly occupy the forefront of a resilient and low-carbon energy future, guiding humanity towards a brighter and more sustainable tomorrow [28-30].

## 2. Existing Structural Problems Regarding the Development and Application of Renewable Energies in China

China, with its swiftly expanding economy and substantial energy requirements, has acknowledged the pressing necessity to shift towards cleaner and more sustainable energy alternatives. Renewable energies, encompassing solar, wind, hydro, and biomass, have emerged as pivotal elements in China's energy strategy to combat climate change and diminish reliance on fossil fuels [31-33]. Despite commendable strides in the adoption of renewable energy technologies, the country confronts several prevailing structural issues that hinder the realization of the full potential of renewable energy development and implementation [34]. This article delves into some of the prominent hurdles that China grapples with in its pursuit of eco-friendlier and sustainable energy future [35].

- **Uneven Regional Distribution:** One of the remarkable structural challenges in China's renewable energy sector is the uneven geographical distribution of resources and developmental initiatives. Renewable energy sources, including wind and solar, are found in greater abundance in specific regions, resulting in disparities in renewable energy generation and utilization throughout the nation. For instance, provinces in the northern and western regions possess substantial wind potential, whereas the southern regions exhibit higher solar potential. The absence of a harmonized and equitable strategy for renewable energy development may impede the overall efficiency and effectiveness of China's transition towards renewable energy [36-40].

- **Integration Challenges:** Effectively integrating renewable energy into the current power grid poses a significant hurdle for China. Renewable energy sources, such as solar and wind, exhibit intermittent characteristics, influenced by weather conditions and daily fluctuations. The grid's constrained adaptability to absorb and regulate variable renewable energy output may lead to curtailing and squandering precious clean energy. Enhancing and modernizing the grid infrastructure to accommodate renewable energy sources in a more flexible and dependable manner are indispensable for surmounting this integration obstacle [41].
- **Policy and Regulatory Ambiguities:** The intricate policy and regulatory landscape concerning renewable energy in China can generate uncertainties for investors and developers. Frequent alterations in policies, subsidy schemes, and feed-in tariffs may dissuade long-term investments in the renewable energy industry. A transparent and steady regulatory framework is crucial to instill confidence and entice private investment in renewable energy ventures [42].
- **Financial Constraints:** Despite substantial investments in renewable energy projects, China encounters difficulties in financing smaller-scale renewable ventures and private enterprises. Access to cost-effective capital and financial incentives can be pivotal in encouraging wider acceptance of renewable energy technologies, particularly in rural and remote regions [43].
- **Land and Environmental Considerations:** Expanding large-scale renewable energy projects, such as solar farms and wind parks, typically demands significant land utilization. Striking a balance between the imperative to expand renewable energy and safeguarding the environment and land usage is a sensitive undertaking. Occasionally, disputes regarding land acquisition and environmental consequences may lead to delays or suspension of renewable energy initiatives [44].
- **Technological Advancements and Innovation:** China has achieved noteworthy advancements in embracing and implementing established renewable energy technologies. Nevertheless, persistent technological breakthroughs and innovation are pivotal for ensuring the sustainable expansion of the renewable energy domain. Devoting resources to research and development in critical domains, such as energy storage, smart grid technologies, and advanced solar and wind technologies, is indispensable for surmounting prevailing constraints and enhancing the overall efficacy and efficacy of renewable energy systems [45].
- **Skills and Workforce Development:** The swift growth of the renewable energy sector in China necessitates a proficient and well-informed workforce. Ensuring the availability of a sufficiently trained and competent workforce to design, install, operate, and maintain renewable energy projects is crucial for attaining enduring sustainability and dependability [46-48].
- **Interplay with Fossil Fuel Industries:** China's shift to renewable energy is intricately connected to its fossil fuel industries. While developing renewable energy is crucial for addressing climate change, it can have implications for workers and communities dependent on fossil fuel extraction and processing. Effectively managing the interaction between renewable energy expansion and the transition of the existing workforce and industries is a complex undertaking that demands meticulous planning and assistance [49-50].

Notwithstanding these current structural issues, China remains dedicated to its renewable energy objectives. The nation's authorities have set ambitious targets for adopting renewable energy and reducing carbon emissions, showcasing their resolve to construct eco-friendlier and sustainable future [51]. Overcoming the identified challenges will demand a collaborative endeavor involving policymakers, industry participants, and local communities to establish a conducive atmosphere for renewable energy development and implementation [52-53]. With appropriate strategies and solutions, China has the capability to emerge as a global frontrunner in renewable energy innovation and set an inspiring precedent for other countries striving to attain a sustainable and low-carbon future [54]. Figure 1 shows the existing structural problems regarding the development and application of renewable energies in China.

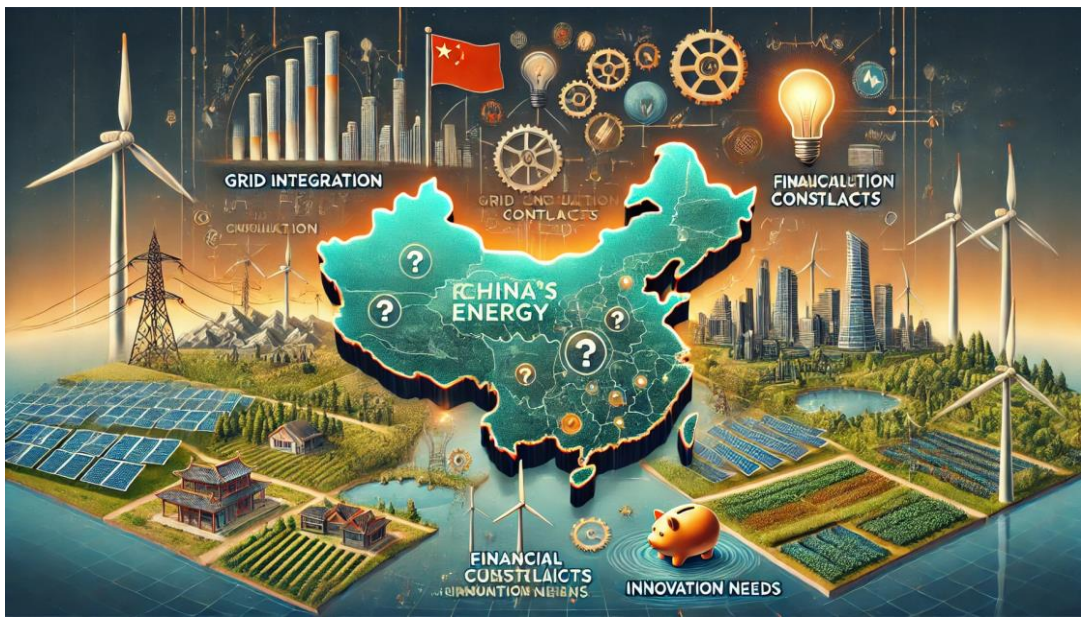

Figure 1. Existing structural problems regarding the development and application of renewable energies in China.

## 3. Drivers and Barriers of the Renewable Energy Industry in China

As the globe confronts the ever-increasing challenges posed by climate change and the imperative for sustainable energy solutions, renewable energy emerges as a pivotal catalyst in curbing greenhouse gas emissions and paving the way for a cleaner future. China, with its vast population and rapidly expanding economy, holds a central position in the worldwide endeavor to shift towards renewable energy sources. Over the last few decades, China has made remarkable progress in the development and implementation of renewable energy technologies [55]. Nonetheless, as with any transformative undertaking, the renewable energy sector in China encounters a combination of drivers that propel its growth and barriers that impede its full potential. This article delves into the key drivers and barriers that are shaping the renewable energy landscape in China [56].

### 3.1. Drivers of the Renewable Energy Industry in China

- **Energy Security and Diversification:** China's substantial dependence on fossil fuels, particularly coal, has underscored the significance of diversifying its energy sources to ensure energy security. By investing in renewable energy technologies, China can diminish its reliance on imported fossil fuels, bolster energy security, and alleviate geopolitical risks connected to fuel imports.
- **Environmental Concerns and Climate Change Mitigation:** China confronts substantial environmental issues, encompassing air pollution and climate change [19]. As the planet's foremost emitter of greenhouse gases, China acknowledges the pressing need to curb its carbon footprint and tackle environmental deterioration. Renewable energy, being carbon-neutral and eco-friendly, serves as a pivotal motivator in China's endeavors to combat climate change and enhance air quality.
- **Economic Growth and Job Creation:** The renewable energy sector offers a pathway for economic expansion and job generation in China. Investments in renewable energy ventures generate employment opportunities in manufacturing, installation, maintenance, and research and development domains, bolstering the growth of an eco-friendly economy.
- **Technological Advancements and Innovation:** China has emerged as a prominent world leader in renewable energy innovations, encompassing solar photovoltaics (PV), wind power, and electric vehicles. Government backing for research and development endeavors has fostered technological progress, rendering renewable energy solutions increasingly cost-efficient and high-performing.
- **Government Policies and Incentives:** China's central and regional governments have implemented visionary policies and incentive programs to bolster the growth of renewable energy [16]. These initiatives comprise feed-in tariffs, tax advantages, and subsidies that allure investments and expedite the adoption of renewable energy technologies.
- **International Cooperation and Commitments:** As a party to global accords such as the Paris Agreement, China remains steadfast in its commitment to diminishing carbon emissions and shifting towards a low-carbon economy. International collaboration and alliances with other nations and organizations have bolstered China's determination to expedite the uptake of renewable energy.

### 3.2. Barriers to the Renewable Energy Industry in China:

- **Intermittency and Grid Integration Challenges:** Renewable energy sources, such as wind and solar, are intermittent and contingent on weather conditions. Incorporating fluctuating renewable energy into the current power grid presents technical obstacles concerning grid stability, balancing, and transmission.
- **Infrastructure Constraints:** Enhancing and updating the grid infrastructure to accommodate a greater portion of renewable energy necessitates significant investments and time. Transmission bottlenecks and insufficient grid capacity in certain areas may impede the effective dissemination of renewable energy [20].

- **Subsidy Reductions and Policy Uncertainty:** Regular alterations in renewable energy incentive programs and regulations can generate uncertainties for investors and developers. Decreases in subsidies can influence the profitability of renewable energy projects and discourage long-term investments.
- **Competing with Fossil Fuels:** Notwithstanding considerable advancement, renewable energy still encounters strong rivalry from firmly established fossil fuel sectors, especially coal. The ongoing dependence on coal-fired power plants in China presents difficulties for the extensive acceptance of renewable energy.
- **Land and Environmental Concerns:** Expanding large-scale renewable energy projects frequently necessitates considerable land utilization [14]. Striking a balance between the necessity for renewable energy growth and environmental conservation and land-use considerations presents a multifaceted undertaking.
- **Financial and Funding Barriers:** Availability of cost-effective capital and extended financing is vital for the enlargement of the renewable energy sector. Smaller-scale initiatives and private enterprises might encounter hurdles in obtaining sufficient funding.
- **Skill Shortages and Workforce Development:** Swift expansion in the renewable energy domain necessitates a proficient and qualified workforce. Guaranteeing a trained and well-informed workforce to design, manage, and uphold renewable energy projects is imperative for the industry's enduring growth.
- **Competing Priorities:** In a nation with a diverse array of priorities, renewable energy might encounter competition from other vital sectors such as infrastructure advancement and poverty reduction.

Despite these obstacles, China's dedication to renewable energy remains steadfast. The resolute approach of the Chinese government to tackle these hurdles through forward-thinking policies, technological progress, and international partnerships emphasizes its commitment to constructing a more ecologically sound and sustainable future. As China persistently channels investments into and advocates for renewable energy, its accomplishments serve as a beacon of inspiration to the global community, uniting in the common pursuit of a low-carbon, resilient, and sustainable planet [11]. Through proactive measures and collective endeavors, China has the capacity to overcome these challenges and chart a course towards a cleaner and more prosperous future fueled by renewable energy. Figure 2 shows the drivers and barriers of the renewable energy industry in China.

4. **Categorization of Proposed Policies Related to the Development of Renewable Energies in China**

China, being the world's most populous nation and a major global energy consumer, confronts substantial challenges in ensuring energy security, fostering economic growth, and addressing environmental concerns. In recent times, the Chinese government has demonstrated a growing focus on the advancement and application of renewable energy sources, recognizing their critical role in the national energy strategy. To expedite the shift towards a more sustainable energy landscape, China has proactively introduced a diverse array of policies designed to facilitate the proliferation of renewable energies. This article presents a comprehensive classification of the proposed policies pertaining to the growth of renewable energies in China, highlighting principal approaches and initiatives that seek to propel the nation's renewable energy metamorphosis.

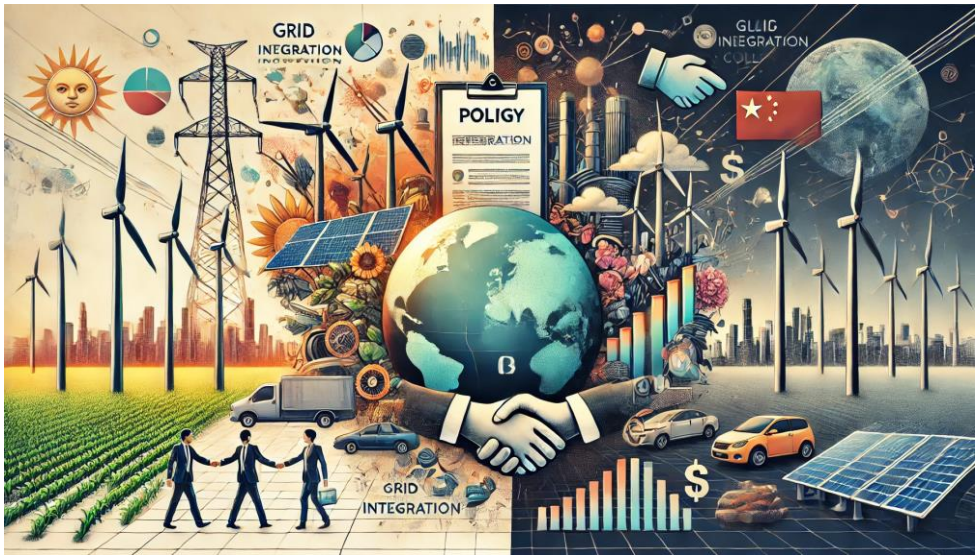

Figure 2: Drivers and barriers of the renewable energy industry in China.

- **Feed-in Tariffs (FiTs) and Subsidies:** Power Purchase Agreements (PPAs) and financial incentives are among the most significant policies in encouraging renewable energy in China. PPAs secure a predetermined price for electricity produced from renewable sources, ensuring stability and enticing investments in renewable energy ventures [21]. Financial incentives, conversely, diminish initial expenditures and render renewable energy technologies more financially feasible for developers and investors. China has provided substantial PPAs and financial incentives for wind, solar, and biomass undertakings to allure private investments and foster the advancement of the renewable energy domain.
- **Renewable Portfolio Standards (RPS) and Targets:** To guarantee a methodical and ambitious enlargement of renewable energy, China has instituted Renewable Portfolio Standards (RPS) and objectives. These policies decree a specific proportion of renewable energy generation in the aggregate energy blend, encouraging the incorporation of renewable sources into the national energy grid. By defining explicit objectives, the Chinese government offers a steady market indication for project developers and investors, nurturing a conducive milieu for renewable energy expansion and market infiltration.
- **Green Bonds and Financing Mechanisms:** To ensure the essential funding for renewable energy projects, China has been proactively advocating green finance initiatives, including the issuance of green bonds. Green bonds are financial instruments explicitly designated to fund environmentally sustainable projects, such as renewable energy ventures. These financing mechanisms allure private and institutional investors interested in supporting sustainable and climate-conscious investments, offering supplementary resources for the proliferation of renewable energies [17].
- **Research and Development (R&D) Incentives:** To stimulate technological progress and innovation in renewable energy, the Chinese government has allocated considerable resources for research and development initiatives. R&D incentives foster the advancement of state-of-the-art renewable energy technologies, improve their efficiency and cost-effectiveness, and bolster China's standing as a worldwide pioneer in renewable energy innovation.
- **Grid Integration and Energy Storage Initiatives:** To tackle the issue of intermittency related to specific renewable energy sources, China is actively pursuing grid integration and energy storage initiatives. Improving and modernizing the national power grid to accommodate the variable nature of renewables is vital for optimizing the use of renewable energy. Moreover, the government is investing in energy storage technologies, like battery storage and pumped hydro storage, to store surplus renewable energy for utilization during periods of low generation.
- **Tax Incentives and Exemptions:** China offers tax benefits and exemptions to incentivize investments in renewable energy ventures. These may involve tax credits for investments in renewable energy, reduced value-added tax (VAT) rates for renewable equipment, and property tax exemptions for renewable energy installations.
- **Demonstration Projects and Pilot Programs:** Demonstration projects and pilot programs have a crucial role in highlighting the feasibility and scalability of renewable energy technologies. China has initiated numerous large-scale demonstration projects in solar, wind, and hydro energy, offering valuable insights and practical experience to enhance renewable energy deployment strategies [19].
- **Public Procurement and Government Support:** China's government assumes a substantial role as a significant energy consumer. By enacting green procurement policies, the government commits to procuring a specific portion of renewable energy for its operations. Government support also encompasses facilitating access to public lands for renewable energy projects and streamlining administrative procedures to accelerate project approvals.
- **International Cooperation and Partnerships:** As a world leader in the deployment of renewable energy, China proactively engages in cooperation with other countries and international organizations to promote renewable energies. These collaborations enable technology transfer, exchange of knowledge, and reciprocal support in the development of renewable energy projects.
- **Energy Efficiency Standards and Regulations:** In addition to renewable energy policies, China places a strong emphasis on energy efficiency standards and regulations to lower overall energy demand. The adoption of energy-efficient appliances, buildings, and industrial processes plays a crucial role in optimizing energy consumption and reducing dependence on energy-intensive sources.
- **Skills Development and Workforce Training:** A competent and knowledgeable workforce is essential for the effective implementation and maintenance of renewable energy projects. China allocates resources to enhance skills development and conduct workforce training programs to equip professionals with the required expertise for the renewable energy sector [14].
- **Community Engagement and Local Benefits:** Involving local communities in the planning and decision-making aspects of renewable energy projects is vital for social acceptance and long-term sustainability. Policies that guarantee local communities benefit from renewable energy initiatives, such as job opportunities and revenue sharing, cultivate a sense of ownership and endorsement for renewable energy development.

- **Competitive Bidding:** In recent years, China has implemented competitive bidding mechanisms for large-scale renewable energy projects. Through these competitive auctions, developers present bids with the most competitive proposed tariffs, and the projects with the lowest prices are chosen. This approach seeks to reduce the expenses of renewable energy and optimize the efficiency of project development.
- **Market-Based Mechanisms:** China is also investigating market-driven mechanisms like carbon pricing and emissions trading as part of its initiatives to advance renewable energies and address climate change. These mechanisms offer economic incentives to curb greenhouse gas emissions and encourage the adoption of low-carbon technologies, including renewable energies.

China's classification of suggested strategies concerning the progression of renewable energies illustrates its all-encompassing and tactical method to encourage the adoption of renewable energy and realize a sustainable energy future. Monetary incentives, regulatory frameworks, research and development backing, and international cooperation have established China as a leading force in the implementation of renewable energy. With a strong emphasis on innovation, sustainability, and community engagement, China is ready to continue its transformative journey towards an energy landscape that is low in carbon and resilient [15]. As technologies progress, costs decrease, and public awareness expands, the combined endeavors of policymakers, businesses, and citizens will undoubtedly shape a world powered by clean and renewable energy sources.

5. **Introduction of Specific Laws and Policies in the Field of Renewable Energies in China**

In recent times, the increasing recognition of ecological concerns and the pressing requirement to tackle climate change have prompted a worldwide shift towards renewable energies. As one of the largest energy consumers and greenhouse gas emitters globally, China holds a pivotal position in the journey towards a sustainable and low-carbon energy future. To realize its ambitious environmental and energy objectives, the Chinese government has introduced a series of targeted laws and policies designed to encourage and expedite the development and adoption of renewable energies [5][6]. This article delves into the key laws and policies that have been implemented in China's renewable energy sector and examines their influence on shaping the nation's renewable energy landscape.

- **Renewable Energy Law (2005):** The Renewable Energy Law, passed in 2005, forms the cornerstone of China's policies regarding renewable energies. This legislation provides the essential legal structure for the advancement, utilization, and administration of renewable energy resources. It mandates power grid enterprises to give preference to procuring electricity generated from renewable sources and mandates that grid connections should be made available to renewable energy projects. Additionally, the law outlines regulations for feed-in tariffs, which serve as a crucial financial incentive for producers of renewable energy.
- **Feed-in Tariffs (FiTs):** Renewable energy feed-in tariffs (FiTs) represent one of the foremost policy tools employed to encourage the growth of renewable energy ventures in China. Through the FiTs system, producers of renewable energy receive assured, fixed prices for the electricity they generate and supply to the grid. These prices are typically higher than those for electricity generated using traditional fossil fuels, rendering renewable energy projects financially appealing to investors. FiTs have demonstrated significant success in propelling the expansion of wind and solar power projects in China.
- **Renewable Portfolio Standards (RPS):** Renewable Portfolio Standards (RPS) serve as another pivotal policy tool harnessed by China to accelerate the deployment of renewable energy. RPS mandates specific, obligatory targets for the proportion of renewable energy within the overall energy mix. These targets vary according to regional contexts and are legally binding on power generation companies, necessitating them to acquire a designated percentage of their electricity from renewable sources [7]. The implementation of RPS creates a consistent and predictable market demand for renewable energy, instigating an impetus for investments in the renewable energy sector.
- **Wind Power Concession Policy:** To foster the growth of wind power projects, China initiated the Wind Power Concession Policy in 2003. This policy provides wind power developers with land and usage rights through a competitive bidding system. The concession period usually spans 25 years, guaranteeing a stable and foreseeable investment environment for wind power projects. This approach encourages greater participation from developers and enhances the deployment of wind power in the country.
- **Solar Power Subsidies:** The Chinese authorities have offered diverse subsidies and economic stimuli to encourage the growth of solar power projects. These subsidies comprise cash incentives, tax breaks, and low-interest loans for solar energy initiatives. The government also incentivizes the incorporation of solar power in buildings by providing financial aid for solar installations on rooftops and facades. These measures aim to accelerate the adoption and deployment of solar energy technologies in the country and foster a more sustainable and cleaner energy landscape.
- **Energy Storage and Grid Integration:** As renewable energy sources, like wind and solar, exhibit intermittent characteristics, China has prioritized energy storage and grid integration to ensure the stability and dependability of the electricity grid. The authorities have implemented policies to bolster the advancement of energy storage technologies, including battery storage and pumped hydro storage, and encourage their seamless integration with the power grid [8]. These initiatives are aimed at enhancing the efficiency and effectiveness of renewable energy utilization while ensuring a more reliable and resilient energy system.
- **Research and Development (R&D) Incentives:** China attaches considerable importance to research and development in the renewable energy domain. The authorities extend financial incentives and assistance to facilitate R&D endeavors, aiming to propel technological breakthroughs and elevate the efficiency of renewable energy technologies. These initiatives play a pivotal role in solidifying China's status as a frontrunner in renewable energy innovation on the world stage.
- **Green Bonds and Green Finance:** Green finance has garnered considerable attention in China as a viable means to finance renewable energy projects. The issuance of green bonds has emerged as a prominent financial mechanism to raise funds for environmentally friendly initiatives, encompassing renewable energy ventures. In addition, China has instituted green development funds and implemented green credit policies to bolster sustainable investments in the renewable energy sector. These initiatives demonstrate the country's commitment to fostering a green and sustainable financial ecosystem to support the growth of renewable energies.
- **Promotion of Electric Vehicles (EVs):** Besides advancing renewable energy generation, China has actively implemented policies to stimulate the widespread adoption of electric vehicles (EVs) as a measure to curtail greenhouse gas emissions in the transportation sector. Various incentives, such as subsidies, tax exemptions, and other favorable measures, have been instituted to bolster the production and purchase of EVs [6]. By promoting the adoption of electric vehicles, China aims to further enhance its efforts in achieving a more sustainable and low-carbon transportation landscape, in line with its environmental and energy goals.
- **International Cooperation and Climate Commitments:** Being a signatory to crucial international climate agreements, such as the Paris Agreement, China has made notable commitments to tackle climate change head-on. In line with its obligations, the Chinese government plays an active role in international cooperation and partnerships concerning renewable energy and clean technologies. By participating in these initiatives, China contributes significantly to the global endeavor to combat climate change and promote a sustainable, low-carbon future. Through collaborative efforts and collective action on a global scale, countries, including China, work together to address the urgent and complex challenges posed by climate change, ensuring a better and more sustainable world for future generations.

The commencement of specific statutes and regulations in the realm of sustainable energies has been pivotal in shaping China's shift towards a sustainable and environmentally friendly energy future. Feed-in premiums, green portfolio standards, wind power concession policies, solar power subsidies, and initiatives for energy reservoir and grid integration have played critical roles in propelling renewable energy implementation in the nation. Furthermore, research and development motivations, eco-friendly finance mechanisms, and promotion of electric vehicles demonstrate China's all-encompassing approach to addressing climate change and fostering eco-friendly progress. China's exertions to advance sustainable energies have positioned the nation as a worldwide leader in the sustainable energy sector [12]. As the world encounters the trials of climate change, China's dedication to sustainable energy progress will persistently have far-reaching consequences for global endeavors to counteract environmental deterioration and construct a more sustainable world. Through well-planned policy frameworks and worldwide collaboration, China is primed to play a central role in shaping the future of sustainable energies on a global scale.

6. **Comparison between China, Canada and the USA towards policy makings for development of renewable energy**

China, Canada, and the USA are three major participants in the worldwide endeavors to encourage the advancement of renewable energy. Each nation has embraced distinct policy methods and tactics to stimulate and bolster the growth of renewable energy. Let's examine the policy formulations for the advancement of renewable energy in these three countries:

- **Renewable Energy Targets and Commitments**

    a. **China:** China has set ambitious renewable energy targets, aiming to increase the share of non-fossil fuels in its energy mix and reach peak carbon dioxide emissions by 2030. The country has committed to achieving 20% of its primary energy consumption from non-fossil fuels by 2030.
    b. **Canada:** Canada has also established renewable energy targets, with a commitment to producing 90% of its electricity from non-emitting sources by 2030. Additionally, several provinces have set individual targets for renewable energy deployment.
    c. **USA:** The USA has implemented a mix of state and federal renewable energy targets. Although there is no national target, many states have set their renewable energy goals, and some cities have committed to achieving 100% renewable energy.
- **Feed-in Tariffs and Renewable Portfolio Standards (RPS)**
    a. **China:** China has implemented feed-in tariffs (FiTs) for various renewable energy sources, guaranteeing fixed electricity prices for renewable energy producers. The country also has a Renewable Portfolio Standard (RPS) system that requires power generation companies to procure a certain percentage of their electricity from renewable sources.
    b. **Canada:** Some provinces in Canada have introduced FiTs for renewable energy producers, while others have adopted RPS policies. These mechanisms vary across provinces and territories, leading to a patchwork of renewable energy incentives.
    c. **USA:** In the USA, renewable energy policies differ among states. Some states have FiTs, while others have RPS requirements for utilities to obtain a portion of their electricity from renewable sources.
- **Incentives and Subsidies**
    a. **China:** China provides significant financial incentives and subsidies to promote the development of renewable energy projects. These include cash grants, tax incentives, and low-interest loans for renewable energy ventures.
    b. **Canada:** Canada also offers financial incentives and subsidies for renewable energy projects. These incentives may include grants, tax credits, and accelerated depreciation for renewable energy investments.
    c. **USA:** The USA provides various federal and state-level incentives for renewable energy, such as Investment Tax Credits (ITC) and Production Tax Credits (PTC), which vary depending on the renewable energy source and location.
- **Research and Development (R&D) Funding**
    a. **China:** China allocates substantial funding for research and development in the renewable energy sector, driving technological advancements and innovations in solar, wind, and other clean energy technologies.
    b. **Canada:** Canada invests in renewable energy R&D through government initiatives and partnerships with universities and research institutions to advance clean energy technologies.
    c. **USA:** The USA also dedicates significant funding to renewable energy R&D through federal agencies and programs aimed at promoting innovation and technological breakthroughs.

- **International Cooperation and Climate Commitments**
    a. **China:** As a signatory to the Paris Agreement, China has committed to take actions to combat climate change and transition to a low-carbon economy. China actively participates in international cooperation and partnerships related to renewable energy and climate change.
    b. **Canada:** Canada is also a party to the Paris Agreement and has made commitments to reduce greenhouse gas emissions. The country collaborates with other nations on renewable energy initiatives and climate action.
    c. **USA:** While the USA withdrew from the Paris Agreement under the previous administration, it has rejoined the agreement, reaffirming its commitment to address climate change and promote renewable energy.

Each nation possesses distinct policy formulations to facilitate the progress of renewable energy. China's centralized methodology, coupled with ambitious objectives and substantial financial backing, has resulted in notable expansion of renewable energy capacity. Conversely, Canada and the USA exhibit more diversified renewable energy policies, encompassing a combination of federal and state/provincial endeavors that differ in scale and aspiration. As the worldwide energy scenario persists in its evolution, the renewable energy policies of these countries will hold crucial significance in shaping their respective energy prospects and contributing to the broader global drive to combat climate change [12][13].

7. **Enhancing Policy Making for Renewable Energies in China: Recommendations for a Sustainable Future**

China's commendable progress in renewable energy development has positioned it as a frontrunner in global renewable energy capacity. Nevertheless, the pursuit of a sustainable and low-carbon future necessitates an ongoing enhancement of policy-making. This article sets forth a series of suggestions to refine the policy-making process for renewable energies in China. The goal is to tackle challenges, optimize current policies, and promote the seamless integration and utilization of renewable energy technologies nationwide.

- **Establish Long-Term Renewable Energy Targets:** One of the primary measures to enhance policy-making for renewable energies in China involves defining clear and long-term objectives. Establishing ambitious yet feasible targets provides a sense of direction and motivation for increased investments in the renewable energy sector. Policymakers should carefully outline targets for individual renewable energy sources, including wind, solar, hydro, and biomass, while also specifying realistic timelines for their implementation. Transparent and quantifiable targets can incentivize project developers, investors, and stakeholders to align their strategies with the national renewable energy objectives, fostering a more cohesive and concerted approach to achieving sustainable energy goals.
- **Promote Grid Integration and Energy Storage:** The remarkable expansion of renewable energy capacity in China has indeed presented challenges in terms of grid integration and management. As a priority, policymakers should concentrate on advancing the development of modern and adaptable power grids capable of effectively handling the fluctuations in renewable energy generation. By providing incentives for the adoption of energy storage technologies, such as battery storage and pumped hydro storage, the intermittency of renewable sources can be effectively mitigated, ensuring a steady and reliable energy supply. Incorporating smart grid technologies and advanced energy storage systems will further enhance the overall efficiency and dependability of the renewable energy sector, facilitating its seamless integration into the existing power grid.

- **Enhance Financial Incentives and Subsidies:** Sustained support through financial incentives and subsidies remains paramount in stimulating private investment and engagement in renewable energy endeavors. Policymakers should carefully consider customized financial mechanisms, such as feed-in tariffs, tax credits, and low-interest loans, to entice private capital into the sector. Ensuring clear and stable policies concerning financial incentives will instill investors with the certainty required for committing to long-term investments in renewable energy projects. This consistency will create a conducive environment for private sector involvement, driving further growth and innovation in the renewable energy industry.
- **Facilitate Research and Development:** To foster ingenuity in renewable energy technologies, policymakers ought to dedicate resources to bolster research and development (R&D) endeavors. Establishing research centers and promoting collaborations between academia, industry, and government will expedite the evolution of state-of-the-art renewable energy solutions. Additionally, implementing a robust intellectual property framework will incentivize companies to invest in R&D and safeguard their innovations, thus cultivating a competitive renewable energy market. Encouraging technology advancements and safeguarding intellectual property rights will cultivate an environment that attracts private sector participation, elevating China's position as a global leader in renewable energy innovation.
- **Strengthen Environmental Regulations:** Rigorous environmental regulations are imperative to guarantee that renewable energy projects uphold sustainable practices [1]. Policymakers should institute unequivocal environmental standards and conduct meticulous assessments of the environmental impact for all renewable energy initiatives. By imposing stringent environmental guidelines, China can uphold its dedication to sustainability while embracing the expansion of renewable energies. Ensuring that renewable energy projects adhere to robust environmental regulations will safeguard natural resources, preserve ecosystems, and mitigate potential negative impacts on the environment, aligning with China's broader commitment to a greener and more sustainable future.
- **Emphasize Community Engagement and Social Acceptance:** Community acceptance is a pivotal aspect for the effective execution of renewable energy initiatives. Policymakers should prioritize community engagement and participation in the planning and decision-making stages [4]. Involving local communities in renewable energy projects can address concerns, minimize opposition, and instill a sense of ownership, fostering a harmonious project development process. Policymakers should establish mechanisms for open dialogue and transparent communication with local communities, incorporating their feedback and concerns into the policymaking process. By promoting inclusivity and actively involving affected communities, China can create a conducive environment for renewable energy projects that are socially and culturally compatible, leading to increased public support and successful implementation.

- **Develop a Holistic Renewable Energy Roadmap:** Developing an all-encompassing renewable energy blueprint that aligns with wider energy and climate policies is vital. This roadmap should merge renewable energy aspirations with energy efficiency strategies, emission abatement goals, and other environmentally friendly development aims. By embracing an integrated approach, China can ensure a seamless transition towards eco-friendlier and sustainable energy terrain.
- **Strengthen International Collaboration:** Global cooperation and knowledge exchange can expedite progress in renewable energy innovations. Decision-makers ought to promote alliances with other nations, institutions, and specialists to share exemplary practices, experiences, and technological proficiency. Drawing inspiration from successful models applied in other regions can inspire inventive policies and strategies customized to China's distinctive circumstances.
- **Implement Clear and Transparent Permitting Processes:** Simplifying the authorization procedures for renewable energy initiatives can lessen delays and uncertainties. Decision-makers ought to give priority to setting up explicit, transparent, and time-limited protocols for obtaining permits and approvals. A predictable authorization process will ease project development and allure more investors to the renewable energy sector.
- **Enhance Renewable Energy Education and Awareness:** Increasing public consciousness and advancing renewable energy education can gain public backing for renewable energy endeavors. Decision-makers should allocate resources to public awareness drives, educational initiatives, and workshops to educate citizens about the advantages of renewable energies [6]. A well-informed and supportive public can create more significant demand for renewable energy and prompt policymakers to prioritize its advancement.

By implementing the suggestions stated in this article, China can enhance its policy-making for the successful incorporation and utilization of renewable energies. Overcoming obstacles, refining current policies, and promoting the efficient advancement of renewable energy technologies can be achieved through these recommendations. With a well-defined and ambitious vision for renewable energy, supported by innovative and sustainable policies, China can pave the way for a more ecologically friendly, cleaner, and robust future. Collaborating with stakeholders, international allies, and local communities, China can take the lead in the global journey towards a sustainable and low-carbon energy landscape.

## 8. Conclusion

In summary, this extensive investigation explores the complex legal facets encompassing the ineffectiveness in the advancement and utilization of sustainable energies, with a specific emphasis on China and illustrative nations. As the international society progressively acknowledges the urgent requirement for eco-friendly alternatives to traditional fossil fuels, eco-friendly power origins have rightly earned considerable consideration as encouraging resolutions to counteract climate change and cultivate ecological preservation. Nonetheless, notwithstanding their transformative capacity, a plethora of legal and regulatory hindrances have emerged, presenting daunting challenges to their smooth incorporation into the energy panorama.

With a resolute and concentrated goal, this investigation has initiated a thorough examination and assessment of the intricate legal structures, regulations, and institutional setups existing in China and chosen countries. The overarching purpose has been to recognize and comprehend potential obstacles and shortcomings that might obstruct the advancement of renewable energy projects and endeavors. By means of a meticulous and comprehensive scrutiny, this research strives to illuminate the precise hindrances impeding the successful execution of renewable energy initiatives.

Embracing a comparative methodology, this investigation has unraveled the varied consequences of different legal systems on the progress and utilization of renewable energies in each examined nation. This nuanced comprehension offers invaluable perceptions into how divergent legal landscapes can either nurture or impede the expansion of renewable energy sectors. Moreover, the scrutiny extends to appraising the efficacy of diverse incentives, subsidies, and environmental regulations that hold a crucial role in advancing and facilitating the widespread uptake of renewable energy technologies.

Furthermore, this inquiry has gone the extra mile in investigating enlightening case studies and extracting exemplary approaches from thoughtfully selected nations that have triumphed over legal hindrances in renewable energy development. These case studies have furnished priceless insights and significant lessons that can steer policymakers, lawmakers, and industry stakeholders in devising effective strategies to navigate the intricate and dynamic legal terrain pertaining to renewable energies.

The consequences of the discoveries from this investigation carry vast potential for the worldwide progression of renewable energy adoption. By identifying focal points for enhancement and acknowledging the critical components that facilitate prosperous integration, this study aspires to assume a crucial role in shaping and influencing the shift towards a more sustainable and environmentally friendly future.

As societies across the globe confront the urgent necessity to transform their energy landscapes, the results of this research hold the potential to forge a path towards a greener, cleaner, and more resilient world. The urgency to adopt renewable energies in response to climate change has never been more critical. The thorough examination of legal barriers and regulatory challenges in this study empowers decision-makers and stakeholders with the knowledge and insights required to cultivate an environment conducive to the development of renewable energy.

By promoting a deeper understanding of the legal landscape, this research seeks to drive progressive policy reforms, support effective governance, and inspire innovative approaches to accelerate the adoption and integration of renewable energies globally. As the urgency to combat climate change intensifies, the transformative potential of renewable energies cannot be underestimated. This study's contributions are envisioned to act as a catalyst for a greener and more sustainable world, where renewable energies lead the way toward a resilient and low-carbon energy future. By bridging the gap between legal intricacies and renewable energy development, this research aspires to spark positive change and contribute to collective efforts aimed at creating a better, cleaner, and more sustainable tomorrow.

As we navigate the challenges ahead, this study envisions a world where renewable energies stand resolute as beacons of hope, guiding humanity towards a brighter and more sustainable future. Through ongoing research, collaboration, and strategic policymaking, the vision of a low-carbon and environmentally conscious energy landscape can be transformed into reality, ensuring a more secure and prosperous future for generations to come. The pursuit of renewable energies remains a shared responsibility, and this study represents one step towards a more sustainable and resilient global energy future. By joining forces and embracing renewable energies, we can collectively shape a world that thrives on clean, renewable, and eco-friendly sources, leaving behind a lasting legacy of positive change for the benefit of our planet and all its inhabitants.

## 9. References

[1] Kavousi-Fard, Abdollah, Morteza Dabbaghjamanesh, Mina Jafari, Mahmud Fotuhi-Firuzabad, Zhao Yang Dong, and Tao Jin. "Digital Twin for mitigating solar energy resources challenges: A Perspective Review." Solar Energy 274 (2024): 112561.

[2] Wang, Boyu, Morteza Dabbaghjamanesh, Abdollah Kavousi-Fard, and Yuntao Yue. "AI-enhanced multi-stage learning-to-learning approach for secure smart cities load management in IoT networks." Ad Hoc Networks 164 (2024): 103628.

[3] Buddy Buruku, "Chinese RE law: China approves renewable energy law in record time", Refocus, May–June 2005

[4] Baohong Jiang, Muhammad Yousaf Raza, "Research on China's renewable energy policies under the dual carbon goals: A political discourse analysis", Energy Strategy Reviews, 1 July 2023

[5] Razmjoui, Pouyan, Abdollah Kavousi-Fard, Tao Jin, Morteza Dabbaghjamanesh, Mazaher Karimi, and Alireza Jolfaei. "A blockchain-based mutual authentication method to secure the electric vehicles' TPMS." IEEE Transactions on Industrial Informatics 20, no. 1 (2023): 158-168.

[6] Mohammadi, Hossein, Shiva Jokar, Mojtaba Mohammadi, Abdollah Kavousi Fard, Morteza Dabbaghjamanesh, and Mazaher Karimi. "A Deep Learning-to-learning Based Control system for renewable microgrids." IET Renewable Power Generation (2023).

[7] Shuzhi Zhang, Guangxiong Xie, "Promoting green investment for renewable energy sources in China: Case study from autoregressive distributed Lagged in error correction approach", Renewable Energy, 10 June 2023

[8] Lu Qiao, Weijia Dong, Xin Lv, "The heterogeneous impacts of M&As on renewable energy firms' innovation: Comparative analysis of China, the US and EU", International Review of Economics & Finance, 5 May 2023

[9] Esapour, Khodakhast, Farid Moazzen, Mazaher Karimi, Morteza Dabbaghjamanesh, and Abdollah Kavousi-Fard. "A novel energy management framework incorporating multi-carrier energy hub for smart city." IET Generation, Transmission & Distribution 17, no. 3 (2023): 655-666.

[10] Haojie Liao, Yuqiang Chen, Hongmei Yang, "Can natural resource rent, technological innovation, renewable energy, and financial development ease China's environmental pollution

burden? New evidence from the nonlinear-autoregressive distributive lag model", Resources Policy, 3 June 2023

[11] Daniel Balsalobre Lorente, Foday Joof, Turgut Türsoy, "Renewable energy, economic complexity and biodiversity risk: New insights from China", Environmental and Sustainability Indicators, 13 March 2023

[12] Tahmasebi, Dorna, Morteza Sheikh, Morteza Dabbaghjamanesh, Tao Jin, Abdollah Kavousi-Fard, and Mazaher Karimi. "A security-preserving framework for sustainable distributed energy transition: Case of smart city." Renewable Energy Focus 51 (2024): 100631.

[13] Tong Su, Yufang Chen, Boqiang Lin, "Uncovering the role of renewable energy innovation in China's low carbon transition: Evidence from total-factor carbon productivity", Environmental Impact Assessment Review, 15 April 2023

[14] Wang, Boyu, Morteza Dabbaghjamanesh, Abdollah Kavousi-Fard, and Shahab Mehraeen. "Cybersecurity enhancement of power trading within the networked microgrids based on blockchain and directed acyclic graph approach." IEEE Transactions on Industry Applications 55, no. 6 (2019): 7300-7309.

[15] Chien-Chiang Lee, Fuhao Wang, Yu-Fang Chang, "Does green finance promote renewable energy? Evidence from China", Resources Policy, 7 March 2023

[16] Jafari, Mina, Abdollah Kavousi-Fard, Morteza Dabbaghjamanesh, and Mazaher Karimi. "A survey on deep learning role in distribution automation system: a new collaborative Learning-to-Learning (L2L) concept." IEEE Access 10 (2022): 81220-81238.

[17] Jianlong Guo, Lifeng Wu, Yali Mu, "An optimized grey model for predicting non-renewable energy consumption in China", Heliyon, 8 June 2023

[18] Mohammadi, Mojtaba, Abdollah Kavousi-Fard, Morteza Dabbaghjamanesh, Amir Farughian, and Abbas Khosravi. "Effective management of energy internet in renewable hybrid microgrids: A secured data driven resilient architecture." IEEE Transactions on Industrial Informatics 18, no. 3 (2021): 1896-1904.

[19] Ashkaboosi, Maryam, Seyed Mehdi Nourani, Peyman Khazaei, Morteza Dabbaghjamanesh, and Amirhossein Moeini. "An optimization technique based on profit of investment and market clearing in wind power systems." American Journal of Electrical and Electronic Engineering 4, no. 3 (2016): 85-91.

[20] ChangZheng Li, Muhammad Umair, "Does green finance development goals affect renewable energy in China", Renewable Energy, 24 December 2022

[21] Dabbaghjamanesh, Morteza, Abdollah Kavousi-Fard, and Zhao Yang Dong. "A novel distributed cloud-fog based framework for energy management of networked microgrids." IEEE Transactions on Power Systems 35, no. 4 (2020): 2847-2862.

[22] Yuanyuan Hao, Xiangdong Li, Muntasir Murshed, "Role of environmental regulation and renewable energy technology innovation in carbon neutrality: A sustainable investigation from China", Energy Strategy Reviews, 23 June 2023

[23] Dabbaghjamanesh, Morteza, Abdollah Kavousi-Fard, and Jie Zhang. "Stochastic modeling and integration of plug-in hybrid electric vehicles in reconfigurable microgrids with deep learning-based forecasting." IEEE Transactions on Intelligent Transportation Systems 22, no. 7 (2020): 4394-4403.

[24] Dabbaghjamanesh, Morteza, Amirhossein Moeini, Nikos D. Hatziargyriou, and Jie Zhang. "Deep learning-based real-time switching of hybrid AC/DC transmission networks." IEEE Transactions on Smart Grid 12, no. 3 (2020): 2331-2342.

[25] Tajalli, Seyede Zahra, Mohammad Mardaneh, Elaheh Taherian-Fard, Afshin Izadian, Abdollah Kavousi-Fard, Morteza Dabbaghjamanesh, and Taher Niknam. "DoS-resilient distributed optimal scheduling in a fog supporting IIoT-based smart microgrid." IEEE Transactions on Industry Applications 56, no. 3 (2020): 2968-2977.

[26] Gámez, María Rodríguez, Antonio Vázquez Pérez, Antonio Sarmiento Será, and Zoila Millet Ronquillo. "Renewable energy sources and local development." International Journal of Social Sciences and Humanities (IJSSH) 1, no. 2 (2017): 10-19.

[27] Del Rio, Pablo, and Mercedes Burguillo. "An empirical analysis of the impact of renewable energy deployment on local sustainability." Renewable and sustainable energy reviews 13, no. 6-7 (2009): 1314-1325.

[28] Dabbaghjamanesh, Morteza, Boyu Wang, Abdollah Kavousi-Fard, Shahab Mehraeen, Nikos D. Hatziargyriou, Dimitris N. Trakas, and Farzad Ferdowsi. "A novel two-stage multi-layer constrained spectral clustering strategy for intentional islanding of power grids." IEEE Transactions on Power Delivery 35, no. 2 (2019): 560-570.

[29] Mohammadi, Hossein, Shiva Jokar, Mojtaba Mohammadi, Abdollah Kavousifard, and Morteza Dabbaghjamanesh. "Ai-based optimal scheduling of renewable ac microgrids with bidirectional lstm-based wind power forecasting." arXiv preprint arXiv:2208.04156 (2022).

[30] Mohammadi, M., Kavousi-Fard, A., Dehghani, M., Karimi, M., Loia, V., Alhelou, H.H. and Siano, P., 2022. Reinforcing Data Integrity in Renewable Hybrid AC-DC Microgrids from Social-Economic Perspectives. ACM Transactions on Sensor Networks.

[31] Papari, B., et al., Effective energy management of hybrid AC–DC microgrids with storage devices. IEEE transactions on smart grid, 2017. 10(1): p. 193-203.

[32] Bornapour, M., et al., Probabilistic optimal coordinated planning of molten carbonate fuel cell-CHP and renewable energy sources in microgrids considering hydrogen storage with point estimate method. Energy Conversion and Management, 2020. 206: p. 112495.

[33] Asl, Dariush Keihan, Ali Reza Seifi, Mohammad Rastegar, Morteza Dabbaghjamanesh, and Nikos D. Hatziargyriou. "Distributed two-level energy scheduling of networked regional integrated energy systems." IEEE Systems Journal 16, no. 4 (2022): 5433-5444.

[34] Baziar, A. and A. Kavousi-Fard, Considering uncertainty in the optimal energy management of renewable micro-grids including storage devices. Renewable Energy, 2013. 59: p. 158- 166.

[35] Razmjouei, Pouyan, Abdollah Kavousi-Fard, Morteza Dabbaghjamanesh, Tao Jin, and Wencong Su. "DAG-based smart contract for dynamic 6G wireless EVs charging

system." IEEE Transactions on Green Communications and Networking 6, no. 3 (2022): 1459-1467.

[36] Zhou, Xuesong, Tie Guo, and Youjie Ma. "An overview on microgrid technology." In 2015 IEEE international conference on mechatronics and automation (ICMA), pp. 76-81. IEEE, 2015.

[37] Olivares, Daniel E., Ali Mehrizi-Sani, Amir H. Etemadi, Claudio A. Cañizares, Reza Iravani, Mehrdad Kazerani, Amir H. Hajimiragha et al. "Trends in microgrid control." IEEE Transactions on smart grid 5, no. 4 (2014): 1905-1919.

[38] Kazemi, Behzad, Abdollah Kavousi-Fard, Morteza Dabbaghjamanesh, and Mazaher Karimi. "IoT-enabled operation of multi energy hubs considering electric vehicles and demand response." IEEE Transactions on Intelligent Transportation Systems 24, no. 2 (2022): 2668-2676.

[39] Kavousi-Fard, Abdollah, Saeed Nikkhah, Motahareh Pourbehzadi, Morteza Dabbaghjamanesh, and Amir Farughian. "IoT-based data-driven fault allocation in microgrids using advanced μPMUs." Ad Hoc Networks 119 (2021): 102520.

[40] Dabbaghjamanesh, M., A. Moeini, M. Ashkaboosi, P. Khazaei, and K. Mirzapalangi. "High performance control of grid connected cascaded H-Bridge active rectifier based on type II-fuzzy logic controller with low frequency modulation technique." International Journal of Electrical and Computer Engineering 6, no. 2 (2016): 484.

[41] Khazaei, P., S. M. Modares, M. Dabbaghjamanesh, M. Almousa, and A. Moeini. "A high efficiency DC/DC boost converter for photovoltaic applications." International Journal of Soft Computing and Engineering (IJSCE) 6, no. 2 (2016): 2231-2307.

[42] Craig, Michael T., Stuart Cohen, Jordan Macknick, Caroline Draxl, Omar J. Guerra, Manajit Sengupta, Sue Ellen Haupt, Bri-Mathias Hodge, and Carlo Brancucci. "A review of the potential impacts of climate change on bulk power system planning and operations in the United States." Renewable and Sustainable Energy Reviews 98 (2018): 255-267.

[43] Sun, Wei, Chen-Ching Liu, and Li Zhang. "Optimal generator start-up strategy for bulk power system restoration." IEEE Transactions on Power Systems 26, no. 3 (2010): 1357-1366.

[44] Khazaei, Peyman, Morteza Dabbaghjamanesh, Ali Kalantarzadeh, and Hasan Mousavi. "Applying the modified TLBO algorithm to solve the unit commitment problem." In 2016 World Automation Congress (WAC), pp. 1-6. IEEE, 2016.

[45] Dabbaghjamanesh, Morteza, Shahab Mehraeen, Abdollah Kavousi-Fard, and Farzad Ferdowsi. "A new efficient stochastic energy management technique for interconnected AC microgrids." In 2018 IEEE Power & Energy Society General Meeting (PESGM), pp. 1-5. IEEE, 2018.

[46] Taherzadeh, Erfan, Shahram Javadi, and Morteza Dabbaghjamanesh. "New optimal power management strategy for series plug-in hybrid electric vehicles." International Journal of Automotive Technology 19 (2018): 1061-1069.

[47] Sabo, Aliyu, Noor Izzri Abdul Wahab, Mohammad Lutfi Othman, Mai Zurwatul Ahlam Mohd Jaffar, and Hamzeh Beiranvand. "Optimal design of power system stabilizer for

multimachine power system using farmland fertility algorithm." International Transactions on Electrical Energy Systems 30, no. 12 (2020): e12657.

[48] Tajalli, Seyede Zahra, Seyed Ali Mohammad Tajalli, Abdollah Kavousi-Fard, Taher Niknam, Morteza Dabbaghjamanesh, and Shahab Mehraeen. "A secure distributed cloud-fog based framework for economic operation of microgrids." In 2019 IEEE Texas Power and Energy Conference (TPEC), pp. 1-6. IEEE, 2019.

[49] Dabbaghjamanesh, Morteza, Boyu Wang, Shahab Mehraeen, Jie Zhang, and Abdollah Kavousi-Fard. "Networked microgrid security and privacy enhancement by the blockchain-enabled Internet of Things approach." In 2019 IEEE Green Technologies Conference (GreenTech), pp. 1-5. IEEE, 2019.

[50] Amrr, Syed Muhammad, MS Jamil Asghar, Imtiaz Ashraf, and Mohammad Meraj. "A comprehensive review of power flow controllers in interconnected power system networks." IEEE Access 8 (2020): 18036-18063.

[51] Al-Muhaini, Mohammad, and Gerald T. Heydt. "Evaluating future power distribution system reliability including distributed generation." IEEE transactions on power delivery 28, no. 4 (2013): 2264-2272.

[52] Ferdowsi, Farzad, Morteza Dabbaghjamanesh, Shahab Mehraeen, and Mohammad Rastegar. "Optimal scheduling of reconfigurable hybrid AC/DC microgrid under DLR security constraint." In 2019 IEEE Green Technologies Conference (GreenTech), pp. 1-5. IEEE, 2019.

[53] Fard, Abdollah Kavousi, Boyu Wang, Omid Avatefipour, Morteza Dabbaghjamanesh, Ramin Sahba, and Amin Sahba. "Superconducting fault current limiter allocation in reconfigurable smart grids." In 2019 3rd International Conference on Smart Grid and Smart Cities (ICSGSC), pp. 76-80. IEEE, 2019.

[54] Hmad, Jihed, Azeddine Houari, Allal El Moubarek Bouzid, Abdelhakim Saim, and Hafedh Trabelsi. "A review on mode transition strategies between grid-connected and standalone operation of voltage source inverters-based microgrids." Energies 16, no. 13 (2023): 5062.

[55] Dehbozorgi, Mohammad Reza, Mohammad Rastegar, and Morteza Dabbaghjamanesh. "Decision tree-based classifiers for root-cause detection of equipment-related distribution power system outages." IET Generation, Transmission & Distribution 14, no. 24 (2020): 5809-5815.

[56] Gong, Helin, Sibo Cheng, Zhang Chen, and Qing Li. "Data-enabled physics-informed machine learning for reduced-order modeling digital twin: application to nuclear reactor physics." Nuclear Science and Engineering 196, no. 6 (2022): 668-693.